\documentclass[12pt]{iopart}
\begin{document}

\title{Torsion and the gravitational interaction}

\author{H I Arcos\footnote[2]{Permanent address: Physics Department, Universidad
Tecnol\'ogica de Pereira, A.A. 97, La Julita, Pereira, Colombia} and  J G Pereira}
\vskip0.5cm
\address{Instituto de F\'{\i}sica Te\'orica, Universidade Estadual
Paulista, Rua Pamplona 145, 01405-900\, S\~ao Paulo SP, Brazil}
\eads{\mailto{hiarcos@ift.unesp.br} and
\mailto{jpereira@ift.unesp.br}}

\begin{abstract}
By using a nonholonomous-frame formulation of the general covariance principle, seen as an
{\em active} version of the strong equivalence principle, an analysis of the gravitational
coupling prescription in the presence of curvature and torsion is made. The coupling
prescription implied by this principle is found to be always equivalent with that of
general relativity, a result that reinforces the completeness of this theory, as
well as the teleparallel point of view according to which torsion does not represent
additional degrees of freedom for gravity, but simply an alternative way of representing
the gravitational field.
\end{abstract}



\section{Introduction}

In general relativity torsion is assumed to vanish from the very beginning. In teleparallel
gravity\footnote{The name teleparallel gravity is commonly used to denote the general
three-parameter theory \cite{obs}. Here, however, we use it as a synonymous for the
teleparallel equivalent of general relativity, a theory obtained for a specific choice of
these parameters.}, on the other hand, instead of torsion, curvature is assumed to vanish.
In spite of this fundamental difference, the two theories are found to yield equivalent
descriptions of the gravitational interaction \cite{equiv}. An immediate implication of
this equivalence is that curvature and torsion might be simply alternative ways of
describing the gravitational field, and consequently related to the same degrees of
freedom of gravity. This property is corroborated by the fact that the symmetric
energy-momentum tensor appears as source in both theories: as the source of curvature in
general relativity, and as the source of torsion in teleparallel gravity. On the other
hand, more general gravity theories, like for example Einstein-Cartan and gauge theories for
the Poincar\'e and the affine groups \cite{gauge}, consider curvature and torsion as
representing independent degrees of freedom. In these models, differently from teleparallel
gravity, torsion becomes relevant only when spins are important \cite{gh96}. This situation
could be achieved either at the microscopic level or near a neutron star, for example.
According to this point of view, torsion might represent additional degrees of freedom in
relation to curvature, and consequently new physics phenomena might be associated with it.
Now, the above described difference rises a conceptual question on the actual role played by
torsion in the description of the gravitational interaction. The basic purpose of this paper
will be to provide a possible answer to this problem.

The approach we are going to use consists in studying the gravitational coupling prescription
in the presence of curvature and torsion, independently of the theory governing the dynamics of
the gravitational field. This can clearly be done because the equation of motion of test
particles, as well as the field equation of any matter field coupled to gravity, can be
obtained independently of the gravitational theory. One has only to consider a gravitational
field presenting curvature and torsion, and use it to obtain, from an independent variational
principle, the particle or field equations in the presence of gravity. Since our interest
here will be to study these particle and field equations---or equivalently, the coupling
prescription of particles and fields to gravitation---we can simply consider this general
gravitational field without specifying the gravitational theory from which it is obtained as
solution.

Our main problem, therefore, will be to obtain the gravitational coupling pre\-scrip\-tion
in the presence of curvature and torsion. This, however, is not an easy task. The
basic difficulty is that, differently from all other interactions of nature, where the
requirement of covariance does determine the gauge connection, in the presence of curvature
and torsion, covariance---seen as a consequence of the strong equivalence principle
\cite{weinberg}---is not able to determine the form of the gravitational coupling
prescription. The reason for this indefiniteness is that the space of Lorentz connections is
an affine space \cite{KoNo}, and consequently one can always add a tensor to a given
connection without destroying the covariance of the theory. As a result of this
indefiniteness, there will exist infinitely many possibilities for the gravitational
coupling prescription. Notice that in the specific cases of general relativity and
teleparallel gravity, characterized respectively by a vanishing torsion and a vanishing
curvature, the above indefiniteness is absent since in these cases the connections are
uniquely determined---and the corresponding coupling prescriptions completely specified---by
the combined use of covariance and the strong equivalence principle. Notice furthermore that
in the case of internal (Yang-Mills) gauge theories, where the concept of torsion is
absent\footnote{ We remark that absence of torsion, which happens in internal gauge
theories, is different from the presence of a vanishing torsion, which happens in general
relativity \cite{differ}.}, the above indefiniteness is not present either.

To deal with the above problem, we are going to use a strategy based on the equivalence
principle. We begin by noticing that, due to the intrinsic relation of gravitation with
spacetime, there is a deep relationship between covariance (either under general coordinate
or local Lorentz transformations) and the strong equivalence principle. In fact, an
alternative version of this principle is the so called {\em principle of general covariance}
\cite{weinberg}. It states that a special relativity equation can be made to hold in the
presence of gravitation if it is generally covariant, that is, it preserves its form under a
general transformation of the spacetime coordinates. The process runs as follows. In order
to make an equation generally covariant, a connection is always necessary, which is in
principle concerned only with the {\em inertial} properties of the coordinate system under
consideration. Then, by using the equivalence between inertial and gravitational effects,
instead of representing inertial properties, this connection can equivalently be assumed to
represent a {\em true gravitational field}. In this way, equations valid in the presence of
gravitation are obtained from the corresponding special relativity equations. Of course, in
a locally inertial coordinate system, they must go back to the corresponding equations of
special relativity. The principle of general covariance, therefore, can be seen as an
{\em active} version of the equivalence principle in the sense that, by making a special
relativity equation covariant, and by using the strong equivalence principle, it is possible
to obtain its form in the presence of gravitation. The usual form of the equivalence
principle, on the other hand, can be interpreted as its {\em passive} version in the sense
that the special relativity equation must be recovered in a locally inertial coordinate
system, where the effects of gravitation are absent. It should be emphasized that general
covariance by itself is empty of physical content as any equation can be {\it made}
generally covariant. Only when use is made of the strong equivalence principle, and the
inertial compensating term is interpreted as representing a true gravitational field, the
principle of general covariance can be seen as an alternative version of the strong
equivalence principle \cite{sciama}.

The above description of the general covariance principle refers to its usual {\it
holonomic} version. An alternative, more general version of the principle can be
obtained in the context of nonholonomic frames. The basic difference  between these
two versions is that, instead of requiring that an equation be covariant under a
general transformation of the spacetime coordinates, in the nonholonomic-frame version
the equation is required to transform covariantly under a {\em local} Lorentz rotation
of the frame. Of course, in spite of the different nature of the involved
transformations, the physical content of both approaches are the same \cite{lorentz}. The
frame version, however, is more general in the sense that, contrary to the coordinate
version, it holds for integer as well as for half-integer spin fields.

The crucial point now is to observe that, when the {\em purely inertial} connection is
replaced by a connection representing a {\em true gravitational field}, the principle of
general covariance naturally defines a covariant derivative, and consequently also a
gravitational coupling prescription. For the cases of general relativity and teleparallel
gravity, the nonholonomic-frame version of this principle has already been seen to yield the
usual coupling prescriptions of these theories \cite{mospe}. The basic purpose of this
paper will then be to determine, in the general case characterized by the simultaneous
presence of curvature and torsion, the form of the gravitational coupling prescription {\em
implied by the general covariance principle}. As we are going to see, in addition to
explaining why there are infinitely many covariant coupling prescriptions for gravitation,
all of them physically equivalent, the resulting coupling prescription will also provide an
alternative interpretation for the role played by torsion in the description of the
gravitational interaction. We begin by reviewing, in the next section, the
nonholonomic-frame formulation of the general covariance principle.

\section{Nonholonomic general covariance principle}

\subsection{Passage to a general nonholonomic frame}

The usual holonomic general covariance principle is quite well known. Here, our interest
here will be its more general nonholonomic-frame version. Let us consider the Minkowski
spacetime of special relativity, endowed with the Lorentzian metric $\eta$. We are going
to use the Greek alphabet $\mu, \nu, \rho, \dots = 0, 1, 2, 3$ to denote holonomic
spacetime indices, and the Latin alphabet $a, b, c, \dots = 0, 1, 2, 3$ to denote
nonholonomic indices related to each local tangent Minkowski spaces. If $\{x^{\mu}\}$ are
inertial Cartesian coordinates in flat spacetime, the basis of (coordinate) vector fields
$\{\partial_\mu\}$ is then a global orthonormal coordinate basis for the flat spacetime.
The frame $\delta_a=\delta_a{}^\mu\partial_\mu$ can then be thought of as a trivial
(holonomous) tetrad, with components $\delta_a{}^\mu$. Consider now a {\em local},
that is, point-dependent Lorentz transformation $\Lambda_a{}^b=\Lambda_a{}^b(x)$. It
yields the new frame
\begin{equation}
e_{a}=e_{a}{}^{\mu }\partial_\mu,
\label{e_a}
\end{equation}
with components
\begin{equation}
e_{a}{}^{\mu }(x) = \Lambda _{a}{}^{b} \; \delta _{b}{}^{\mu }.
\label{deltabar}
\end{equation}
Notice that, on account of the locality of the Lorentz transformation,
the new frame $e_a$ is nonholonomous,
\begin{equation}
[e_a, e_b] = f_{ab}{}^c \; e_c,
\label{nonholo}
\end{equation}
with $f_{ab}{}^c$ the coefficient of nonholonomy. Now, making use of the orthogonality
property of the tetrads, we see from Eq.~(\ref{deltabar}) that the Lorentz group
element can be written in the form
$\Lambda_b{}^d = e_b{}^\rho \delta_\rho{}^d$. Using this expression, it follows that
\begin{equation}
(e_a \Lambda_b{}^d)\Lambda^c{}_d =
\frac{1}{2} \left(f^c{}_{ab} + f^c{}_{ba} - f_{ba}{}^c \right).
\label{deltabar e f}
\end{equation}

On the other hand, the action describing a free particle is
\begin{equation}
S = - m c \int ds,
\label{action}
\end{equation}
with $ds = (\eta_{\mu \nu} \, dx^\mu dx^\nu)^{1/2}$ the spacetime invariant interval. Seen
from the holonomous frame $\delta_a$, the corresponding equation of motion is given by
\begin{equation}
\frac{d v^a}{d s} = 0,
\label{plivre4}
\end{equation}
where $v^a = \delta^a{}_\mu v^\mu$, with $v^\mu = (dx^{\mu}/ds)$ the holonomous
particle four-velocity. Seen from the nonholonomous frame $e_a$, a straightforward
calculation shows that the equation of motion (\ref{plivre4}) is
\begin{equation}
\frac{dV^{c}}{ds} + \frac{1}{2}
\left( f^c{}_{ab} + f^c{}_{ba} - f_{ba}{}^c \right) V^a V^b = 0,
\label{plivre5}
\end{equation}
where $V^c = \Lambda^c{}_d \, v^d = e^c{}_\mu v^\mu$, and use has been made of
Eq.~(\ref{deltabar e f}). It is important to emphasize that, although we are in the
flat spacetime of special relativity, we are free to choose any tetrad $\{e_{a}\}$ as
a moving frame. The fact that, for each $x \in M$, the frame $e_{a}$ can be
arbitrarily rotated introduces the compensating term $\frac{1}{2}\left(f^c{}_{ab} +
f^c{}_{ba} - f_{ba}{}^c \right)$ in the free-particle equation of motion. This term,
therefore, is concerned only with the (inertial) properties of the frame. It is also
important to remark that the equation of motion (\ref{plivre5}) can be obtained from
the action (\ref{action}) provided it is written in the nonholonomous frame, in which
case it assumes the form
\begin{equation}
S = - m c \int (\eta_{ab} \, e^a e^b)^{1/2},
\label{action2}
\end{equation}
where, owing to the fact that the Lorentz transformation is an isometry, the metric
$\eta_{ab} = \Lambda_a{}^c \Lambda_b{}^d \; {\eta}_{cd}$ is kept fixed, and $e^a =
(\Lambda^a{}_d \, \delta^d{}_\mu) dx^\mu$.

\subsection{Using the equivalence between inertial and gravitational effects}

According to the general covariance principle, the equation of motion valid in
the presence of gravitation can be obtained from the corresponding special
relativistic equation by replacing the inertial compensating term by a connection
$\Gamma_{ab}{}^c$ representing a true gravitational field. We consider here only
Lorentz-valued connections, and consequently vanishing nonmetricity. In this case, in
the presence of both curvature and torsion, such a connection satisfies \cite{nonholo}
\begin{equation}
\Gamma_{ba}{}^c - \Gamma_{ab}{}^c = T_{ab}{}^c + f_{ab}{}^c,
\end{equation}
with $T_{ba}{}^c$ the torsion of the connection $\Gamma_{ab}{}^c$. Use of this
equation for three combination of indices gives
\begin{equation}
\Gamma_{ab}{}^c =
\frac{1}{2} \left( f^c{}_{ab} + f^c{}_{ba} - f_{ba}{}^c \right) +
\frac{1}{2} \left( T^c{}_{ab} + T^c{}_{ba} - T_{ba}{}^c \right).
\label{genecom}
\end{equation}
Accordingly, the compensating term of Eq.~(\ref{plivre5}) can be written in the form
\begin{equation}
\frac{1}{2} \left( f^c{}_{ab} + f^c{}_{ba} - f_{ba}{}^c \right) =
\Gamma_{ab}{}^c - K_{ab}{}^c,
\label{equiva}
\end{equation}
where
\begin{equation}
K_{ab}{}^c = \frac{1}{2} (T^c{}_{ab} + T^c{}_{ba} - T_{ba}{}^c)
\end{equation}
is the contortion tensor. Equation~(\ref{equiva}) is actually an expression of the
equivalence principle. In fact, whereas its left-hand side involves only {\it
inertial} properties of the frames, its right-hand side contains purely {\it
gravitational} quantities. Using this expression in Eq.~(\ref{plivre5}), we get
\begin{equation}
\frac{dV^{c}}{ds} + \Gamma_{ab}{}^c V^a V^b = K_{ab}{}^c V^a V^b.
\label{plivre5c}
\end{equation}
This is the particle equation of motion in the presence of curvature and torsion that
follows from the principle of general covariance. It entails a very peculiar
interpretation for contortion, which appears playing the role of a gravitational
force \cite{glf}. Because of the identity
\begin{equation}
\Gamma_{ab}{}^c - K_{ab}{}^c = {\stackrel{\circ}{\Gamma}}_{ab}{}^c,
\label{rela}
\end{equation}
with ${\stackrel{\circ}{\Gamma}}_{ab}{}^c$ the spin connection of general
relativity, the equation of motion (\ref{plivre5c}) is found to be equivalent with the
geodesic equation of general relativity.

\subsection{Gravitational coupling prescription}

The equation of motion (\ref{plivre5c}) can be written in the form
\begin{equation}
V^\mu D_\mu V^c \equiv
V^\mu \left[ \partial_\mu V^c + (\Gamma_{\mu b}{}^c - K_{\mu b}{}^c) V^b \right] = 0,
\label{cova}
\end{equation}
with $D_\mu$ a covariant derivative. Applied on a general vector field $X^{c}$, it
assumes the form
\begin{equation}
D_\mu X^c =
\partial_\mu X^c + (\Gamma_{\mu b}{}^c - K_{\mu b}{}^c) \; X^b.
\end{equation}
Using the vector representation of the Lorentz generators \cite{ramond},
\begin{equation}
(S_{ab})^c{}_d = i (\delta_a{}^c \, \eta_{bd} - \delta_b{}^c \, \eta_{ad}),
\end{equation}
it becomes
\begin{equation}
D_{\mu} X^{c} =
\partial_{\mu} X^{c} - \frac{i}{2} (\Gamma_{\mu}{}^{a b} - K_{\mu}{}^{a b}) \;
(S_{ab})^c{}_d \; X^d.
\label{acomivetor2}
\end{equation}

Now, although obtained in the case of a Lorentz vector field (four-velocity), the
compensating term (\ref{deltabar e f}) can be easily verified to be the same for any
field. In fact, denoting by $g \equiv g(\Lambda)$ the element of the Lorentz group in an
arbitrary representation, it can be shown that
\begin{equation}
(e_{a} g) g^{-1} = - \frac{i}{4}
\left(f_{cab} + f_{cba} - f_{bac} \right) \, J^{bc},
\end{equation}
with $J^{bc}$ denoting the corresponding Lorentz generator. In this case, the covariant
derivative (\ref{acomivetor2}) will have the form
\begin{equation}
D_{\mu} = \partial_{\mu} - \frac{i}{2} \left(\Gamma_{\mu}{}^{a b} -
K_{\mu}{}^{a b} \right) J_{ab}.
\label{genecova}
\end{equation}
Consequently, the coupling prescription---in the presence of curvature and torsion---of
fields carrying an arbitrary representation of the Lorentz group, will be
\begin{equation}
\partial_a \equiv \delta^\mu{}_a \partial_\mu \rightarrow
D_a \equiv e^\mu{}_a D_\mu.
\end{equation}
Of course, due to the relation (\ref{rela}), it is clearly equivalent with the
coupling prescription of general relativity.

\section{Universality of the general relativity prescription}

\subsection{The connection space: characterizing the affinity}

A general connection space is an infinite, homotopically trivial affine
space \cite{singer}. In the specific case of Lorentz connections, a point in this space
will be a connection
\begin{equation}
A = A_\mu{}^{b c} \, J_{b c} \; dx^\mu
\end{equation}
presenting simultaneously curvature and torsion, 2-forms defined respectively by
\begin{equation}
R = dA + A A \equiv \nabla_A A
\label{curve}
\end{equation}
and
\begin{equation}
T = de + A e \equiv \nabla_A e,
\label{torsion}
\end{equation}
where $e = e^a{}_\mu dx^\mu \partial_a$, and $\nabla_A$ denotes the covariant
differential in the connection $A$. Under a {\it local} Lorentz transformation, a
Lorentz connection transforms according to
\begin{equation}
A \rightarrow A^\prime = g A g^{-1} + g d g^{-1}.
\end{equation}
The curvature and torsion 2-forms are covariant under these transformations:
\begin{equation}
R^\prime = g R g^{-1} \quad {\rm and} \quad T^\prime = g T g^{-1}.
\end{equation}

Given two connections $A$ and $\bar{A}$, the difference
\begin{equation}
k = \bar{A} - A
\label{trans1}
\end{equation}
is also a 1-form assuming values in the Lorentz Lie algebra, but transforming
covariantly:
\begin{equation}
k = g k g^{-1}.
\end{equation}
Its covariant derivative is consequently given by
\begin{equation}
\nabla_Ak = dk + \{A, k\}.
\end{equation}
It is then easy to verify that, given two connections such that $\bar{A}=A+k$,
their curvature and torsion will be related by
\begin{equation}
\bar{R} = R + \nabla_A k + k \, k
\end{equation}
and
\begin{equation}
\bar{T} = T + k \, e.
\end{equation}
The effect of adding a covector $k$ to a given connection $A$, therefore, is to
change its curvature and torsion 2-forms \cite{alana}.

Let us now rewrite Eq.~(\ref{trans1}) in components:
\begin{equation}
A_{ab}{}^c \equiv e^\mu{}_a A_{\mu b}{}^c = \bar{A}_{ab}{}^c - k_{ab}{}^c.
\end{equation}
Since $k_{ab}{}^c$ is a Lorentz-valued covector, it is necessarily anti-symmetric in the
{\it last two} indices. We notice that the presence of nonmetricity would spoil the
anti-symmetry in the last two indices, and consequently the connection would not be
Lorentz valued. Separating $k_{ab}{}^c$ in the symmetric and anti-symmetric parts
in the {\it first two} indices, we get
\begin{equation}
k_{ab}{}^c = \frac{1}{2} (k_{ab}{}^c + k_{ba}{}^c) +
\frac{1}{2} (k_{ab}{}^c - k_{ba}{}^c).
\end{equation}
If we call
\begin{equation}
k_{ab}{}^c - k_{ba}{}^c = t_{ab}{}^c,
\end{equation}
we see that $t_{ab}{}^c$ will automatically satisfy $t_{ba}{}^c = - t_{ab}{}^c$.
It is then easy to verify that the symmetric part turns out to be
\begin{equation}
k_{ab}{}^c + k_{ba}{}^c = t^c{}_{ab} + t^c{}_{ba}.
\end{equation}
Therefore, $k_{ab}{}^c$ can always be written in the form
\begin{equation}
k_{ab}{}^c = \frac{1}{2} (t^c{}_{ab} + t^c{}_{ba} - t_{ba}{}^c).
\label{contor}
\end{equation}
This means essentially that the difference between any two Lorentz-valued connections, that
is, the affinity covector, has the form of a contortion tensor.

\subsection{Equivalence under translations in the connection space}

As already discussed, due to the affine character of the connection space, one can
always add a tensor to a given connection without spoiling the covariance of the
derivative (\ref{genecova}). Since adding a tensor to a connection corresponds just
to redefining the origin of the connection space, this means that covariance does not
determine a preferred origin for this space. Let us then analyze the physical meaning of
translations in the connection space. To begin with, we take again the connection
appearing in the covariant derivative (\ref{genecova}), which is given by
\begin{equation}
A_{ab}{}^c \equiv \Gamma_{ab}{}^c - K_{ab}{}^c,
\label{equiva2}
\end{equation}
where we have used the relation $A_{ab}{}^c = e^\mu{}_a A_{\mu b}{}^c$. A translation in the
connection space with parameter $k_{ab}{}^c$ corresponds to
\begin{equation}
\bar{A}_{ab}{}^c = A_{ab}{}^c + k_{ab}{}^c \equiv
\Gamma_{ab}{}^c - K_{ab}{}^c + k_{ab}{}^c.
\label{equiva3}
\end{equation}
Now, since $k_{ab}{}^c$ has always the form of a contortion tensor, as
given by Eq.~(\ref{contor}), the above connection is equivalent to
\begin{equation}
\bar{A}_{ab}{}^c = A_{ab}{}^c - \bar{K}_{ab}{}^c,
\end{equation}
with $\bar{K}_{ab}{}^c = K_{ab}{}^c - k_{ab}{}^c$ another contortion tensor.

Let us then consider a few particular cases. First, we choose  $t_{ba}{}^c$ as the
torsion of the connection $\Gamma_{ab}{}^c$, that is, $t_{ba}{}^c=T_{ba}{}^c$. In
this case, the last two terms of Eq.~(\ref{equiva3}) cancel each other, yielding
$\bar{K}_{ab}{}^c = 0$. This means that the torsion of $\bar{A}_{ab}{}^c$ vanishes,
and we are left with
\begin{equation}
\bar{A}_{ab}{}^c = {\stackrel{\circ}{\Gamma}}{}_{ab}{}^c,
\label{einstein}
\end{equation}
with ${\stackrel{\circ}{\Gamma}}{}_{ab}{}^c$ denoting the torsionless spin connection
of general relativity. On the other hand, if we choose $t_{ba}{}^c$ such that
\begin{equation}
t_{ba}{}^c = T_{ba}{}^c - f_{ba}{}^c,
\end{equation}
the connection $\Gamma_{ab}{}^c$ vanishes, which characterizes tele\-parallel gravity. In
this case, the resulting connection has the form \cite{tsc}
\begin{equation}
\bar{A}_{ab}{}^c = - {\stackrel{w}{K}}{}_{ab}{}^c,
\label{weitzen}
\end{equation}
where ${\stackrel{w}{K}}{}_{ab}{}^c$ is the Weitzenb\"ock contortion, that is, the
contortion tensor written in terms of the Weitzenb\"ock torsion
${\stackrel{w}{T}}{}_{ab}{}^c = - f_{ab}{}^c$. There are actually infinitely many
choices for $t_{ba}{}^c$, each one defining a different translation in the connection
space, and consequently yielding a connection with different curvature and torsion. All
of them, however, are ultimately equivalent with the coupling prescription of general
relativity as for all cases we have the identity
\begin{equation}
\bar{A}_{ab}{}^c = A_{ab}{}^c - \bar{K}_{ab}{}^c \equiv
{\stackrel{\circ}{\Gamma}}{}_{ab}{}^c.
\end{equation}
It is important to emphasize that, despite yielding physically equivalent coupling
prescriptions, the physical equations are not covariant under a translation in the
connection space. For example, under a particular such translation, the geodesic equation
of general relativity becomes the force equation of teleparallel gravity, which are
completely different equations. These two equations, however, as well as any other
obtained through a general translation in the connection space, are equivalent in the
sense that they describe the same physical trajectory.

\section{Final remarks}

The general covariance principle, seen as an active version of the usual (passive) strong
equivalence principle, naturally defines a coupling prescription of any field to
gravitation. By considering the case of a Lorentz connection presenting simultaneously
curvature and torsion, we have shown that the gravitational coupling prescription {\em
implied by the general covariance principle} is such that it preserves the equivalence with
the coupling prescription of general relativity. This result reinforces the completeness of
general relativity, as well as gives support to the point of view of teleparallel gravity,
according to which torsion does not represent additional degrees of freedom of gravity,
but simply an alternative way of representing the gravitational field. As a consequence,
it becomes a matter of convention to describe gravitation by curvature, torsion, or by a
combination of them.

The above result can be better understood if we remember that gravitation presents two
alternative descriptions. In fact, according to the gauge approach provided
by teleparallel gravity, the gravitational interaction is described by a force
equation similar to the Lorentz force equation of electrodynamics, with contortion
playing the role of force. On the other hand, due to the universality of free fall, it
is also possible to describe gravitation, not as a {\em force}, but as a geometric {\em
deformation} of flat Minkowski spacetime. This is the approach used by general
relativity, in which the gravitational field is supposed to produce a {\em curvature} in
spacetime, and its action on (structureless) particles is described by letting them
follow the geodesics of the curved spacetime. In this approach, therefore, geometry
replaces the concept of gravitational force, and the trajectories are determined, not by
force equations, but by geodesics. In the general case, characterized by a connection
presenting both curvature and torsion, the gravitational interaction is described by a
mixture of force and geometry. Since all cases are ultimately equivalent, how much of
the interaction is described by curvature (geometry), and how much is described by
contortion (force), is a matter of convention. All these equivalent cases are related
through translations in the connection space, whose affinity allows the addition of
a tensor (actually, a Lorentz-valued covector) to a given connection without destroying
the covariance of the theory.

Sometimes, the autoparallel curves are considered as describing the motion of a spinless
particle in the presence of curvature and torsion \cite{kleinert}. Such curves are
characterized by parallel transporting the tangent vector itself, and by the fact that they
do not represent the shortest line between two points of spacetime. However, it has already
been shown that autoparallels cannot be obtained from a Lagrangian formalism \cite{ms},
which means that a spinless particle following such a trajectory does not have a Lagrangian.
Taking into account that the energy-momentum is defined as the functional derivative of the
Lagrangian with respect to the metric tensor (or equivalently, to the tetrad field), it
would not be possible to define an energy-momentum density for such particle. Despite the
non-existence of experimental data, the results presented here can be considered as a
theoretical evidence favoring the fact that, even in the presence of curvature and torsion,
a spinless particle will always follow a trajectory that can ultimately be represented by a
geodesic of the underlying Riemannian spacetime.

The new interpretation for torsion here described differs from the usual one used in
theories of the Einstein-Cartan type, for example. It is, however, self-consistent, and
presents several conceptual advantages in relation to the Einstein-Cartan interpretation: it
preserves the role played by torsion in teleparallel gravity; it is consistent with the
general covariance principle, that is, with the strong equivalence principle; the
corresponding gravitational coupling prescription can be applied in the Lagrangian
or in the field equations with the same result; when applied to describe the interaction of
the electromagnetic field with gravitation, it does not violate the U(1) gauge invariance of
Maxwell's theory\footnote{This comes from the equivalence with general relativity, whose
coupling prescription does not violate the U(1) gauge invariance of Maxwell's theory. For the
specific case of teleparallel gravity, see \cite{maxwell}.}. In addition, considering that, at
least up to now, there is no compelling experimental evidence for {\it new physics associated
with torsion}, we could say that our results agree with the available experimental data. For
example, no new gravitational physics has ever been reported near a neutron star, where
torsion would become important according to the Einstein-Cartan type theories. Of course, it
should be remarked that, because of the weakness of the gravitational interaction, no
experimental data exist on the coupling of the spin of the fundamental particles to
gravitation.

From the classical point of view, as we have seen, two connections differing by a
Lorentz-valued covector yield the same physical result. On the other hand, from the point of
view of a prospective connection-based quantum theory for gravitation \cite{ash}, these two
connections might produce different observable effects \cite{amelino}. If, however, the
connection-space translation could somehow be gauged in the quantum theory, the choice of the
connection would then correspond to a kind of ``gauge choice'', and the final quantum theory
should naturally yield the same physical equivalence of the classical theory. The connection
choices (\ref{einstein}) and (\ref{weitzen}) could accordingly be named respectively the
Einstein and the Weitzenb\"ock gauges. Of course, there exist infinitely more physically
equivalent gauge choices, each one characterized by a different proportion between curvature
and torsion, and differing by a translation in the connection space. It should be noted finally
that, if gravitation eventually loses its universal character at the quantum
level\footnote{There are some controversies concerning this point; for a discussion, see
e.g. \cite{global}}, the general relativistic geometrical description (in terms of curvature)
of the gravitational interaction would break down, and due to the fact that teleparallel
gravity does not require the validity of the equivalence principle \cite{wep}, the
Weitzenb\"ock gauge may eventually become mandatory.

\ack
The authors thank R.\ Mosna and R.\ Aldrovandi for useful discussions. They also thank
FAPESP-Brazil, CNPq-Brazil and COL\-CIENCIAS-Colombia for financial support.


\end{document}